\documentstyle[epsf]{article}
\begin{document}
\begin{center}
{\Large {\bf Strong-Coupling Gauge Theory of Nodal Spinons \\
and \\
\vspace{0.2cm}
Antiferromagnetic Long-Range Order
}}

\vskip 1cm

{\Large  Ikuo Ichinose}  \\
{ Department of Applied Physics,
Nagoya Institute of Technology, Nagoya, 466-8555 Japan}

\end{center}

\begin{center} 
\begin{bf}
Abstract
\end{bf}
\end{center}

In this paper we shall study a gauge theory of nodal spinons which appears as 
a low-energy effective theory for antiferromagnetic (AF) Heisenberg models.
In most of studies on the nodal spinons given so far, the gauge interaction
between spinons was assumed weak and nonperturbative effects like
instantons and vortices were ignored.
In the present paper, we shall study strong-coupling gauge theory of nodal
spinons and reveal its nontrivial phase structure.
To this end, we employ recently developed lattice gauge theory
techniques for studying finite-temperature and finite-density gauge theory.
At low temperature and low spinon-density region, 
an AF long-range order exists.
As temperature and/or density of spinons are increases, a phase transition
to nonmagnetic phase takes place.
Order of the phase transition is of second (first) order for low (high)
density region of spinons.
At a quantum critical point at vanishing temperature $T=0$, 
abrupt change of spinon
density occurs as a function of the chemical potential.
Implications of the results to the heavy-fermion materials and the high-$T_c$
cuprates are discussed.

\newpage
\setcounter{footnote}{0} 

\section{Introduction}
Since the discovery of the high-$T_c$ cuprates, search for nontrivial quantum
spin liquid for antiferromagnetic (AF) materials is one of the most 
challenging topics in the condensed matter physics.
Among them, possibility of appearance of spin ${1 \over 2}$ fermionic 
or bosonic quasi-particles called
spinons has been studied from various point of view.
In certain class of the AF Heisenberg models in two dimensions (2D),
by using slave-particle representation for the quantum spin operators and 
mean-field theory (MFT) like approximation of decoupling fields, 
one can simple show 
that spinons with a relativistic dispersion appear as low-energy 
excitations\cite{AM}.
If the fluctuations of the decoupling field are ignored, spin-spin
correlation functions exhibit a power-law decay.
This type of spin states were dubbed algebraic spin liquids 
(ASL)\cite{RW,ASL}.

In the ASL, besides spinons, there exist gapless gauge-field excitations,
which are phase degrees of freedom of the decoupling field and 
its fluctuations may destabilize the MF picture of the ASL.
It is known that the gauge-field effects in 2D is very strong 
and low-energy excitations are often drastically 
influenced by the gauge interactions.
Generally speaking there are two phases in the gauge theory; one is 
confinement phase in which the gauge field fluctuates very strong
and charged particles by themselves cannot exist as low-energy excitations.
The other is deconfinement phase (it is furthermore classified into Coulomb and
Higgs phases) in which nonperturbative effects of the gauge-field 
fluctuations can be neglected and charged particles and gauge boson itself
are low-energy excitations.
In the discussion of the ASL, the deconfinement phase of the U(1) gauge field
is assumed.
Though there is no Maxwell term (or no plaquette term) of the gauge field
in the original action,
this assumption is justified if the number of gapless spinons $N_s$ is very 
large because of the shielding effects of the gapless matters.\footnote{
For discussions on the problem of deconfinement
phenomenon in gauge theories coupled with gapless matter fields, 
see Refs.\cite{CSS1,CSS2,CSS3}}. 
However our recent study on the deconfinement phase transition by a
large number of gapless matter fields indicates that the critical 
value of $N_s$ is rather large.
In Ref.\cite{TIM}, we numerically investigated multi-flavor CP$^1$ models
on a lattice and found that the critical number of two-component complex
bosons for the deconfinement transition is 14 which is larger than the number
of the two-component Dirac fermions appearing in the typical ASL's.
If the gauge theory of nodal spinons is in a confinement phase,
weakly interacting spinons may be a good picture in the distance scale 
of the original lattice spacing but the groundstate and picture
of the quasi-excitations at long distances are drastically changed by 
the violent fluctuations of the gauge field.

In this paper, we shall study the strong-coupling gauge theory of
nodal spinons from the above point of view.
We shall clarify its phase structure and quasi-excitations.
To this end, we employ recently developed lattice gauge-theory techniques
for studying finite-temperature and finite-density gauge 
theory\cite{Nishida}.

This paper is organized as follows.
In Sec.2, we briefly review the ASL in which relativistic Dirac fermions
appear.
In Sec.3, we introduce a lattice model describing strong-coupling
gauge theory of spinons.
Symmetries of the model is also discussed.
In Sec.4, an effective action is derived by integrating over the gauge field
and spinons.
In Sec.5, by using a MFT like approximation we clarified the phase structure
of the system.
Section 6 is devoted for discussion and conclusion.

\setcounter{equation}{0}
\section{Algebraic Spin Liquid}

In this section we shall briefly review the ASL.
Let us consider $S=1/2$ Heisenberg model on the square lattice,
\begin{equation}
H_{\rm AF}=J \sum_{x,i}\vec{S}_x\cdot\vec{S}_{x+i}+\cdots,
\label{HAF}
\end{equation}
where $\vec{S}_x$ is the spin operator at site $x=(x_1,x_2)$, 
$i$ is the direction
index (sometimes it also denotes the unit vector in the $i$-th direction),
and $J>0$ is the nearest-neighbor (NN) exchange coupling.
In the Hamiltonian (\ref{HAF}), there may be other interaction terms between
spins like the next-NN exchanges, and the ring exchanges, etc which are
represented by the ellipsis.
The spin operator $\vec{S}_x$ can be represented by the two-component
fermionic spinon operator $f_x=(f_{x,1}, f_{x,2})=
(f_{x,\uparrow}, f_{x,\downarrow})$ as,
\begin{equation}
\vec{S}_x={1 \over 2}f^\dagger_x\vec{\sigma}f_x,
\label{spinonop}
\end{equation}
where $\vec{\sigma}$ is the Pauli spin matrices and $f_x$
must satisfy the local constraint 
$\sum_{\alpha=1}^2f^\dagger_{x\alpha}f_{x\alpha}=1$ because
the magnitude of the quantum spin is $1/2$.
By substituting Eq.(\ref{spinonop}) into Eq.(\ref{HAF}) and decoupling
the quartic terms of $f_x$ in $H_{\rm AF}$ 
by introducing decoupling fields $D_{xi}$,
the decoupled Hamiltonian is obtained as follows,
\begin{equation}
H_{\rm dc}=-\sum_{xi}D_{xi}f^\dagger_{x+i}f_x+\mbox{H.c.}+|D_{xi}|^2/J 
+\cdots.
\label{HDc}
\end{equation} 
From Eq.(\ref{HDc}), it is obvious that the phase of $D_{xi}$,
$iA_{xi}\equiv\log (D_{xi}/|D_{xi}|)$,
behaves like a U(1) gauge field under a gauge transformation,
\begin{equation}
f_x \rightarrow e^{i\alpha_x}f_x, \;\;
D_{xi} \rightarrow e^{i\alpha_{x+i}} D_{xi}e^{-i\alpha_x}.
\label{Gtrf}
\end{equation}
In the MF approximation, the fluctuations of $D_{xi}$
are totally ignored and $D_{xi}$ is replaced with its expectation
value $D_{xi}^0\equiv \langle D_{xi}\rangle$ in Eq.(\ref{HDc}).

Low-energy excitations are determined by the pattern and symmetry of 
the MF $D_{xi}^0$. 
For example in the 
$\pi$-flux state, phase of the product of four $D_{xi}$'s around the
smallest plaquettes is $\pi$.
In most of general flux states, the low-energy excitations are described by 
two-component (or four-component)
relativistic Dirac fields $\psi^a$ where $a=1,\cdots, n_f$ is the flavor index
(the number of nodes).
For the $\pi$-flux state, $H_{\rm dc}$ in Eq.(\ref{HDc}) is nothing but the 
Susskind lattice fermion and $n_f$ is the number of the species doublers
($n_f=2$).
Restoring the local gauge invariance by $A_{i}$ and imposing local constraint
$f^\dagger_xf_x=1$ by introducing a gauge field $A_{0}$, 
an effective theory for low-energy excitations is given as follows
in the continuum,
\begin{equation}
S_{\rm spinon}=\int d^3x \bar{\psi}\Big[\gamma_\mu(\partial_{\mu}+
iA_\mu)\Big]\psi+\cdots,
\label{S}
\end{equation}
where the notations are standard and the ellipsis in Eq.(\ref{S}) 
represents higher-derivative terms.
We have set $\hbar=1$ and also the ``speed of light" (the speed of 
$\psi$-quanta) $v_s=1$.
The original AF Heisenberg model is the half-filled state of $f_x$
because the number of $f_x$ is unity at each site.
In the effective field theory of the Dirac fermions, {\em this half-filled 
state just corresponds to the Dirac sea} in which all the states 
with negative energy are filled.

Besides the original spin SU(2) symmetry, the continuum action $S_{\rm spinon}$
has the relativistic invariance as well as the flavor symmetry.
It is interesting to see how the original spin operators are
expressed in terms of the Dirac fermions.
From the discussion in Refs.\cite{susskind,ASL},
the N\'eel vector $\vec{N}_x\equiv (-)^{x_1+x_2}\vec{S}_x$  
is related with $\psi$ as 
\begin{equation}
(-)^{x_1+x_2}\vec{S}_x\propto \bar{\psi} \vec{\sigma}\psi,
\label{Neel}
\end{equation} 
where $\vec{\sigma}$ is the Pauli spin matrix acting on the spin index.

If nonperturbative effects of the gauge field $A_\mu$ are negligible and 
the system (\ref{S}) is in a deconfinement phase of the gauge dynamics,
the system can be studied, e.g., by the $1/n_f$ expansion.
Composite operators of $\psi$ may acquire an anomalous dimension,
but the massless Dirac fermions survive as low-energy excitations
at long distances.
However if the system is in a confinement phase,
the structure of the groundstate and low-energy excitations are 
drastically influenced by $A_\mu$.
For 2D case (and probably also for 3D strong-coupling case), 
it is expected that 
for sufficiently large $n_f$ the gauge system is in a deconfinement phase
even for the strong-coupling limit without the Maxwell term,
whereas it is in a confinement phase for small $n_f$.
That is, there exists a critical number of the massless field separating
confinement and deconfinement phases.
Very recently we estimated this critical number $n^c_f$ for massless 
two-component
bosonic systems by investigating multi-flavor CP$^1$ models in $(2+1)$D
and obtained $n^c_f=14$, which is rather large compared to the
spinon number in the typical ASL's.
Then it is interesting to see how the groundstate and the low-energy
excitations of the nodal spinons are changed if the nonperturbative 
effects like instanton condensation occurs and the gauge dynamics is
in a confinement phase.
We address this problem in the following sections.


\setcounter{equation}{0}
\section{Lattice Gauge Theory of Spinons}

\begin{figure}[htbp]
\begin{center}
\epsfxsize=3cm
\epsffile{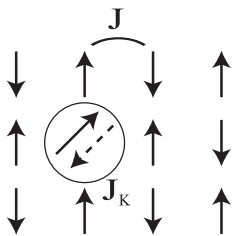}
\caption{Hybridization of the localized spin and the conduction electron
in the AF background of the localized spins. 
Spin-singlet state of $f_x$ (a solid arrow) and
conduction electron (a dashed arrow) appears as a result of the Kondo
coupling $J_K$. 
}
\label{fig1}
\end{center}
\end{figure}

In this section, we shall study the gauge system (\ref{S}) assuming
that the gauge dynamics is in the confinement phase, i.e., $n_f<n_f^c$.
We consider the system at finite-temperature ($T$) and with finite density
of spinons.
Motivation to consider the finite-density system comes from not only
the hole-doped high-$T_c$ cuprates but also from the
physics of the heavy-fermion materials.
The Hamiltonian of a Kondo-Heisenberg model for the heavy fermions is given as,
\begin{equation}
H_{\rm KH}=\sum_k\epsilon_k c^\dagger_{k\alpha}c_{k\alpha}+
J_K\sum_x \vec{S}_x\cdot c^\dagger_x\vec{\sigma}c_x+H_{\rm AF},
\label{HKH}
\end{equation}
where $c_{k\alpha}$ represents the conduction electrons with the energy
$\epsilon_k$, and $J_K$ is the Kondo coupling.
The Kondo coupling prefers hybridization of the conduction electron 
$c_x$ and the localized spin $f_x$, whereas the AF coupling 
tends to make the AF order for $\vec{S}_x$.
In the real experiments, external conditions like an external pressure 
enhances the itinerant properties of the (localized) electrons. 
This fact implies that the external pressure changes density of states
and effectively increases the Kondo coupling 
$J_K$ and as a result the number of electrons $f_x$ making a spin-singlet
pair with the conduction electrons is also increased. (See Fig.\ref{fig1})
In the following discussion, the density of spinons is controlled
by introducing the chemical potential $\mu_s$ instead of considering
the Kondo coupling explicitly.

In order to study the strong-coupling gauge theory for the gauge
system (\ref{S}),
we introduce a $d$-dimensional {\em spacetime} hypercubic lattice.
The lattice sites are denoted again by $x$, but $x$ does not represent 
individual atoms of the original AF Heisenberg model; rather we are 
considering ``coarse grained" lattice model valid at long length scales
where the gauge-field fluctuations are very strong.

Lattice model is obtained by replacing derivative with lattice
difference in the action (\ref{S}),
\begin{equation}
S_\psi={1\over 2}\sum_{x,\mu}\Big[\bar{\psi}_x\gamma_\mu 
U_\mu(x)\psi_{x+\mu}
-\bar{\psi}_{x+\mu}\gamma_\mu U^\dagger_\mu(x)\psi_x\Big], 
\label{Sn}
\end{equation}
where hereafter $\mu=0, \cdots, d$ and $U_\mu(x)$ is the compact U(1) 
gauge field defined on the links $(x,x+\mu)$.
However as well-known in the lattice gauge theory, the action $S_{\psi}$
contains additional low-energy modes besides $k_\mu \sim 0$, the species 
doublers.
In order to decrease these extra low-energy modes and to make the
symmetry of the system clear, we make the following change of variables,
\begin{equation}
\psi_x=T(x)\chi_x, \; \; \bar{\psi}_x=\bar{\chi}_xT^\dagger(x),
\;\; T(x)=(\gamma_0)^{x_0}\cdots (\gamma_d)^{x_{d}}.
\label{chi}
\end{equation}
Substituting Eq.(\ref{chi}) into the action (\ref{Sn}), we obtain
\begin{equation}
S_\psi={1\over 2}\sum_{x,\mu}\Big[\eta_\mu(x)\bar{\chi}_xU_\mu(x)
\chi_{x+\mu}-\eta_\mu(x)\bar{\chi}_{x+\mu}U^\dagger_\mu(x)\chi_x
\Big],
\label{Schi1}
\end{equation}
where $\eta_0(x)=1$ and $\eta_j(x)\equiv (-)^{x_0+\cdots +x_{j-1}}$.
From the action (\ref{Schi1}), it is obvious that the Dirac spinor indices
are {\em diagonal} in the $\chi_x$ representation, and then we can 
decrease the number of the low-energy excitations by reducing the number 
of spinor indices.
Hereafter let us put the ``flavor number" of $\chi_x$ as $N_f$
(i.e., $\chi^a_x; a=1,\cdots, N_f$). 
In the continuum, the above system (\ref{Schi1}) contains 
$2^{[(d+1)/2]}\cdot N_f$ massless Dirac fermions.
Therefore $2^{[(d+1)/2]}\cdot N_f=2n_f=$[spin degrees of freedom]
$\times$[the number of nodes].

As we explained above, we shall consider the system of a finite-density
spinons.
Then the action is given as follows by introducing the chemical
potential $\mu_s$\cite{chemical},
\begin{equation}
S_\chi={1\over 2}\sum_{x,\mu}\Big[r_\mu(x)\bar{\chi}_xU_\mu(x)
\chi_{x+\mu}-r^{-1}_\mu(x)\bar{\chi}_{x+\mu}U^\dagger_\mu(x)\chi_x
\Big],
\label{Schi2}
\end{equation}
where
$r_0(x)=e^{\mu_s}$ and $r_j(x)=\eta_j(x)$.
Because of the particle-hole (electron-positron) symmetry, the physical results
are unchanged under a transformation $\mu_s \rightarrow -\mu_s$. 
We assume $\mu_s<0$ hereafter for the density of spinons decreases
by the hole doping or the Kondo coupling.

Before starting study of the spinon-gauge system,
we have to notice the following things,
\begin{enumerate}
\item As we consider the strong gauge-coupling limit, there are no plaquette
(no Maxwell) terms of $U_\mu(x)$ in the action.
\item As $U_0(x)$ is the Lagrange multiplyer field for the local constraint
on the number of $\chi_x \; (f_x)$, we have to introduce another matter field
representing the ``background charge" to make the total charge vanishing.
In real materials, this degrees of freedom correspond to doped holes or
spin-singlet pairs made of a conduction electron and $\vec{S}_x$.
\end{enumerate}
The requirement 1 in the above is obvious. For 2, we shall introduce the
matter field $\varphi^a_x$ ($a=1,\cdots,N_f$) 
by adding the following terms to the action,
\begin{equation}
S_\varphi={1\over 2}\Big[r'_0\bar{\varphi}_xU_0(x)\varphi_{x+\hat{0}}
-(r'_0)^{-1}\bar{\varphi}_{x+\hat{0}}U^\dagger_0(x)\varphi_x\Big],
\label{Sphi}
\end{equation}
where $r'_0=e^{\mu'_s}$.
From $S_\varphi$, it is obvious that $\varphi_x$ carry the same charge
with $\chi_x$ with respect the Lagrange multiplyer field $U_0(x)$ 
and they have no hopping terms and are always staying 
at site $x$.
For the original AF system with ``hole doping", the slave-particle
representation is useful, in which a hard-core boson operator $b_x$
is introduced for the ``hole state" (or spin-singlet state)
and the physical state is restricted to
$(f^\dagger_xf_x+b^\dagger_xb_x-1)|phys\rangle=0$.
See Fig.\ref{fig1}.
The above $\varphi_x$ on the coarse grained lattice corresponds to 
$b_x$ on the original lattice.\footnote{More precisely, the flavor number 
of $\varphi_x$ sholud be $N_f/2$ for the hole-doped case.
In this sense, we are think of the two-channel Kondo-Heisenberg model
in which two kinds of spin-singlet pair exist.
Qualitative phase structure is the same for the $N_f$ and $N_f/2$ cases.}

In the later calculation of the free energy, we shall see that 
the chemical potentials $\mu_s$ and $\mu'_s$ always appear in the linear
combination $\mu_s-\mu'_s$ as a result of integration over $U_0(x)$.
This means that the summation of densities of $\chi_x$ and $\varphi_x$ 
is automatically vanishing as required by the above physical-state
condition.

From $S_\chi$ (\ref{Schi2}), the gauge system has the following global
U($N_f$)$\times$ U($N_f$) symmetry;
\begin{equation}
\chi_x \rightarrow W_D \chi_x, \;\; 
\bar{\chi}_x \rightarrow \bar{\chi}_x W_D^\dagger, \;\; 
W_D \in \mbox{U($N_f$)},
\label{sym1}
\end{equation}
and for $W_A \in \mbox{U($N_f$)}$,
\begin{eqnarray}
\chi_x &\rightarrow& \left\{
  \begin{array}{ccc}
        W_A \chi_x, & x & \in \mbox{even site}  \\
        W^{-1}_A \chi_x, & x & \in \mbox{odd site} 
  \end{array}
\right.   \nonumber  \\
\bar{\chi}_x &\rightarrow& \left\{
  \begin{array}{ccc}
  \bar{\chi}_xW_A, & x &\in \mbox{even site}  \\
  \bar{\chi}_xW^{-1}_A, & x & \in \mbox{odd site}.
\end{array}
\right.
\label{sym2}
\end{eqnarray}
It is shown that the ``N\'eel order parameter" $\bar{\psi}_x\vec{\sigma}
\psi_x=\bar{\chi}_x\vec{\sigma}\chi_x$ is connected with 
$\bar{\psi}_x\psi_x=\bar{\chi}_x
\chi_x$ by the second symmetry in Eq.(\ref{sym2}), e.g.,
\begin{equation}
\bar{\chi}_x\chi_x \rightarrow \bar{\chi}_xe^{i{\pi \over 4}\sigma_j}
e^{i{\pi \over 4}\sigma_j}\chi_x =i\bar{\chi}_x\sigma_j\chi_x,
\label{sym3}
\end{equation}
where $\sigma_j$'s are the Pauli spin matrices acting on the spin indices
and they are the generators of SU(2)$\subset$U($N_f$).
For the spinon-gauge system in the continuum (\ref{S}), it is shown that
$\bar{\psi}_x\vec{\sigma}\psi_x=\bar{\chi}_x\vec{\sigma}\chi_x$ and 
$\bar{\psi}_x\psi_x=\bar{\chi}_x\chi_x$
has the same scaling dimension to all order of the $1/n_f$ expansion.
In the present lattice model, these operators are 
connected by the explicit symmetry.

In the following section, we shall study the lattice spinon-gauge model
in the strong-coupling limit by using recently developed techniques
of the lattice gauge theory\cite{Nishida}.

\setcounter{equation}{0}
\section{Free Energy of the Lattice Spinon-Gauge System}

The partition function $Z$ is given as,
\begin{equation}
Z=\int [D\bar{\chi}D\chi D\bar{\varphi}D\varphi DU_0DU_j]e^{-S_\chi-S_\varphi},
\label{Z}
\end{equation}
where $j=1,\cdots, d$.
As we are considering the finite-temperature system, we impose the 
(anti)periodic boundary condition on the field variables.
Temperature $T$ is given as $T=(aN_0)^{-1}$ where $a$ is the lattice spacing
and $N_0$ is the number of the temporal sites.
Hereafter we simply set $a=1$.
Unit of energy is $\hbar v_s/a$ and will be set unity.

There are no gauge-plaquette terms because we are think of the 
strong-coupling limit.
Then we can perform the integration over the spatial gauge field $U_j(x)$
($j=1, \cdots, d$); the one-link integral.
We employ the $1/d$-expansion as in the most of calculations and obtain,
\begin{equation}
\int [DU_j]e^{-{1\over 2}\sum_{x,j}[\eta_j(x)\bar{\chi}_xU_j(x)
\chi_{x+j}-
\eta_j(x)\bar{\chi}_{x+j}U^\dagger_j(x)\chi_x]}  
=\exp \Big[\sum_{x,y}\mbox{Tr}(M_xV_M(x,y)M_y)\Big],
\label{onelink2}
\end{equation}
where 
\begin{equation}
M^{ab}_x=\bar{\chi}_x^a\chi^b_x,  \;\; 
V_M(x,y)={1\over 8}\sum_{j=1}^{d}(\delta_{x,y+j}+\delta_{x,y-j}).
\label{VM}
\end{equation}
 
Let us introduce auxiliary fields $\Phi^{ab}_x$ corresponding to 
the composite fields $M^{ab}_x$.
To this end, we use the following identity,
\begin{eqnarray}
 &&\exp \Big[\sum_{x,y}\mbox{Tr}(M_xV_M(x,y)M_y)\Big] \nonumber  \\
&&\;\; =\int [D\Phi]\exp \Big[- \sum_{x,y}\Big\{\mbox{Tr}(\Phi_x
V_M(x,y)\Phi_y)+2 \mbox{Tr}(\Phi_xV_M(x,y)M_y)\Big\}\Big].
\label{Phi}
\end{eqnarray}

In the following, we shall calculate the free energy (the effective potential)
of $\Phi_x$.
Because of the symmetry Eqs.(\ref{sym1}) and (\ref{sym2}), we can simply set
$\Phi_x^{ab}=\Phi_x\delta^{ab}$.
(See for example, Eq.(\ref{sym3}).)
Furthermore we consider homogeneous configuration like $\Phi_x=\Phi$,
because it is the lowest-energy configuration.
Low-energy excitations of $\Phi_x$ will be discussed in Sec.6.
Then, each term in Eq.(\ref{Phi}) becomes as,
\begin{equation}
\sum_{x,y}\mbox{Tr}(\Phi_xV_M(x,y)\Phi_y) \Rightarrow \sum_x
{dN_f \over 4}\Phi^2,
\label{Phi^2}
\end{equation}
and 
\begin{equation}
\sum_{x,y}\mbox{Tr}(\Phi_xV_M(x,y)M_y) 
\Rightarrow {d \over 4} \Phi \sum_x M^{aa}_x 
= {d \over 4} \Phi \sum_x \bar{\chi}_x\chi_x.
\label{chi^2}
\end{equation}
From Eqs.(\ref{Phi^2}) and (\ref{chi^2}), the partition function can be 
written as 
\begin{equation}
Z=\int [DU_0][D\bar{\chi}D\chi][D\bar{\varphi}D\varphi]
e^{-S'[U_0,\chi,\Phi]-S_\varphi},
\label{Z2}
\end{equation}
where 
\begin{equation}
S'[U_0,\chi,\Phi]=\sum_x \Big[{1\over 2}\bar{\chi}_xe^\mu U_0(x)
\chi_{x+\hat{0}}-{1\over 2}\bar{\chi}_{x+\hat{0}}e^{-\mu}
U^\dagger_0(x)\chi_x 
+{dN_f \over 4}\Phi^2-{d\over 2}\Phi \bar{\chi}_x\chi_x\Big].
\label{S'}
\end{equation} 

Next we shall integrate over $\chi_x$ and $\varphi_x$.
As they satisfy the anti-periodic boundary condition in the $0$-th direction,
we introduce the Fourier transformed fields,
\begin{equation}
\chi_{x_0,\vec{x}}={1 \over \sqrt{N_0}}\sum^{N_0}_{n=1}
e^{ip_nx_0}\tilde{\chi}(n,\vec{x}),  \;\;
\bar{\chi}_{x_0,\vec{x}}={1 \over \sqrt{N_0}}\sum^{N_0}_{n=1}
e^{-ip_nx_0}\tilde{\bar{\chi}}(n,\vec{x}),
\label{Ftrf}
\end{equation}
where $p_n=\pi(2n-1)/N_0$ and similarly for $\varphi_x$.
In order to perform the path-integral over the Fourier transformed
fields, we take the Polyakov gauge for $U_0(x)$ as follows,
\begin{equation}
U_0(x_0,\vec{x})=e^{i\phi(\vec{x})/N_0}\;\;\; \mbox{for all $x_0$}.
\label{PG}
\end{equation}
After integration over $\tilde{\chi}(n,\vec{x})$ etc, we have 
\begin{equation}
Z=\int [D\phi]\prod_{\vec{x}}\prod_{n=1}^{N_0/2}
\Big[\sin^2 p_n(\phi)+\Big({d\Phi \over 2}\Big)^2\Big]^{N_f} 
 \Big[\sin^2 p'_n(\phi)\Big]^{N_f}\cdot e^{-{dN_f \over 4}\sum_x\Phi^2},
\label{ZZ}
\end{equation}
where 
\begin{equation}
p_n(\phi)\equiv p_n+{\phi(\vec{x})\over N_0}-i\mu_s, \;\;
p'_n(\phi)\equiv p_n+{\phi(\vec{x})\over N_0}-i\mu'_s.
\label{pp}
\end{equation}

The product $\prod_{n=1}^{N_0/2}$ in Eq.(\ref{ZZ}) can be calculated by
using the usual techniques of the finite-temperature field theory,
\begin{eqnarray}
Z&=& \int [D\phi]\prod_{\vec{x}}\Big[\cosh (N_0E)\cdot 
\cos (\phi(\vec{x})-iN_0\mu'_s) \nonumber  \\
&& \;\;+{1\over 2}\cos(2\phi(\vec{x})-iN_0(\mu_s-\mu'_s))
+{1\over 2}\cosh(N_0(\mu_s-\mu'_s))\Big]^{N_f}\cdot 
e^{-{dN_f \over 4}\sum_x\Phi^2},
\label{ZZZ}
\end{eqnarray}
where 
$
E={\rm arcsinh} \Big({d\Phi \over 2}\Big).
$

Finally we have to integrate over $\phi(\vec{x})$ in Eq.(\ref{ZZZ})
to obtain the free energy $F_{\rm eff}=-\log Z/N_0V_s$, where 
the temperature $T=1/N_0$ and $V_s=\sum_{\vec{x}}$.
Unfortunately, the explicit form of $F_{\rm eff}$ cannot be obtained for
general $N_f$.
For each $N_f$, we integrated over $\phi(\vec{x})$ in Eq.(\ref{ZZZ}) and
then $F_{\rm eff}$ is obtained as follows for $N_f=2$, 
\begin{equation}
F_{\rm eff}={d \over 2}\Phi^2-T \log \Big[{1\over 4}
\cosh^2\Big({\mu_s-\mu'_s \over T}\Big)  
+{1\over 2}\cosh^2\Big({E \over T}\Big)+{1\over 8}\Big].
\label{F2}
\end{equation}
For $N_f=4$,
\begin{eqnarray}
F_{\rm eff}&=&d\Phi^2-T \log \Big[{1\over 16}
\cosh^4\Big({\mu_s-\mu'_s \over T}\Big)
+{15 \over 8}\cosh^2\Big({\mu_s-\mu'_s\over T}\Big)
\cos^2\Big({E\over T}\Big) \nonumber \\
&&+{3\over 8}\cosh^4\Big({E\over T}\Big)+{3\over 32}
\cosh^2\Big({\mu_s-\mu'_s\over T}\Big)+{3\over 2^7}\Big].
\label{F4}
\end{eqnarray}
Similar expression is obtained for other values of $N_f$.
As stated above, the chemical potentials always appear in the combination
$\mu_s-\mu'_s$ and therefore densities of $\chi_x$ and $\varphi_x$ are 
automatically related with each other,
\begin{equation}
\rho_\chi=-{\partial F_{\rm eff} \over \partial \mu_s}
={\partial F_{\rm eff} \over \partial \mu'_s}
=-\rho_\varphi,
\label{rho}
\end{equation}
as required by the original local constraint $f^\dagger_xf_x+b^\dagger_xb_x=1$.
Hereafter we set $\mu_T\equiv-\mu_s+\mu_s'\ge 0$.

In the following section, we shall study the phase structure of the 
spinon-gauge system by using the free energy obtained in this section.
To this end, we shall assume a typical form of $F_{\rm eff}$ instead of
investigating some specific cases separately.


\setcounter{equation}{0}
\section{Phase structure of the spinon-gauge system}

As we show in the previous section, the free energy $F_{\rm eff}$
has a rather complicated form.
In this section we shall study the phase structure of the spinon-gauge
system assuming the following ``phenomenological" free energy,
\begin{equation}
F_{\rm eff}= {dN_f\over 4} \Phi^2-T \log 
\Big[\cosh\Big({N_f\mu_T \over T}\Big)
+B\cosh\Big({N_fE \over T}\Big)+C\Big],
\label{Fef2}
\end{equation}
where $B$ and $C$ are parameters.
For general $N_f$, there appear more complicated terms as in Eq.(\ref{F4}),
but the essential feature of the free energy and the phase structure can be
described by Eq.(\ref{Fef2}).

First let us consider the small $\Phi$ region.
For $\Phi \ll 1$,
\begin{equation}
F_{\rm eff}= {dN_f\over 4} \Phi^2-{BdN_f \over 8[\cosh({N_f\mu_T\over T})
+B+C]}\cdot {dN_f\Phi^2\over T} +O(\Phi^4).
\label{Phi2}
\end{equation}
As long as the coefficient of the $\Phi^4$-term is positive, $F_{\rm eff}$
exhibits the second-order phase transition and the
chemical potential at the critical points $\mu^c_T$ is given as a function 
of $T$,
\begin{equation}
\mu^c_T={T\over N_f}\mbox{arccosh}\Big[{BdN_f \over 2T}-B-C\Big].
\label{mucr}
\end{equation}
In the region $\mu_{T}<\mu^c_T$, the field $\Phi$ has a nonvanishing
expectation value and the symmetry is spontaneously broken.
In the AF spin model, an AF order exists in this region.

In particular from Eq.(\ref{mucr}), the critical $T$ at 
$\mu_T=0$, which corresponds to the pure Heisenberg model without 
hole doping, obtained as 
\begin{equation}
T_N={BdN_f \over 2(1+B+C)},
\label{Neel2}
\end{equation}
which corresponds to the N\'eel temperature.

As $T$ is lowered, the coefficient of $\Phi^4$ becomes negative and 
the phase transition changes from a second-order to first-order one.
We call this point tricritical point.
The tricritical temperature $T_{\rm tri}$ can be easily obtained 
from Eq.(\ref{Fef2}) as
\begin{equation}
T_{\rm tri}={N_f \over 2}.
\label{Tri}
\end{equation}
From Eq.(\ref{mucr}), the chemical potential at the tricritical point is 
given as,
\begin{equation}
\mu_{T}^{\rm tri}={1\over 2}\cdot\mbox{arccosh}\Big((d-1)B-C\Big).
\label{mutri}
\end{equation}
If ${dB\over B+C+1}<1$, the phase transition is everywhere of first order.

Next let us consider the $T=0$ case; the finite-density 
{\em quantum phase transition} (QPT).
In this case, the free energy reduces to,
\begin{equation}
F_{\rm eff}[\Phi]={dN_f \over 4}\Phi^2-N_f\cdot \mbox{max}
\Big\{\mu_T, E[\Phi]\Big\}.
\label{FT0}
\end{equation}
At $\Phi=0$, $E[\Phi=0]=0$ and therefore $F_{\rm eff}[\Phi=0]=-N_f\mu_T$.
It is obvious that $\Phi=0$ is a local minimum of $F_{\rm eff}[\Phi]$.
There is another local minimum at $\Phi_0$, where $\Phi_0$ is the
solution to the equation $\partial F_{\rm eff}[\Phi]/\partial \Phi=0$
for $E[\Phi]>\mu_T$.
$\Phi_0$ is easily obtained as 
\begin{equation}
\Phi^2_0={2\sqrt{1+d^2}-2 \over d^2},
\end{equation}
and the local minimum is calculated as 
\begin{equation}
F_{\rm eff}[\Phi_0]={dN_f \over 4}\Phi_0^2
-N_fE[\Phi_0]<0.
\end{equation}
As $\mu_T$ increases, the global minimum changes from $\Phi_0$
to $\Phi=0$. 
The critical chemical potential is then given by
\begin{equation}
\mu_{T0}^{c}=E[\Phi_0]-{d\over 4}\Phi_0^2,
\label{muT0}
\end{equation}
and the expectation value of $\Phi$ behaves as 
\begin{equation}
\langle \Phi\rangle(T=0)=\left\{
  \begin{array}{cc}
  \Phi_0& \;\; \mbox{for} \; \;\mu_T<\mu_{T0}^c \\
   0 & \;\; \mbox{for} \; \;\mu_T>\mu_{T0}^c.
  \end{array} 
\right.
\label{PhiT0}
\end{equation}
Similarly, the density of the spinons $\rho\equiv \rho_\chi=-\rho_\varphi$ is 
calculated from 
Eq.(\ref{FT0}),
\begin{equation}
\rho={\partial F_{\rm eff} \over \partial \mu_T}=
\left\{
  \begin{array}{cc}
  0 & \;\; \mbox{for} \; \;\mu_T<\mu_{T0}^c \\
  -N_f & \;\; \mbox{for} \; \;\mu_T>\mu_{T0}^c.
  \end{array}
\right.
\label{rhoT0}
\end{equation}

The above result Eq.(\ref{PhiT0}) and (\ref{rhoT0}) indicates that 
a strong first-order phase transition from ordered to disordered phases
takes place at $T=0$, and an abrupt change of the density of the spinons 
from the empty state to the saturated state 
also occurs there.
For $\mu_T>\mu_{T0}^c$, the density of the localized
electrons participating in AF magnetism are vanishing i.e., all electrons are
itinerant.

\begin{figure}[htbp]
\begin{center}
\epsfxsize=6cm
\epsffile{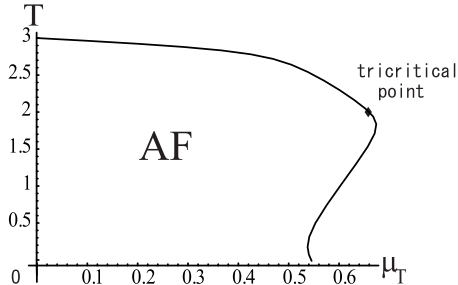}
\caption{Phase diagram obtained from the free energy by using MFT 
like approximation. Low-$T$ and low-density region of spinons has
a AF long-range order. Symbol $\bullet$ denotes the tricritical point
at $\mu_T=\mu^{\rm tri}_T$. Above (Below) $T_{\rm tri}$, the phase 
transition is of second (first) order.
}
\label{figPD}
\end{center}
\end{figure}

\begin{figure}[htbp]
\begin{center}
\epsfxsize=6cm
\epsffile{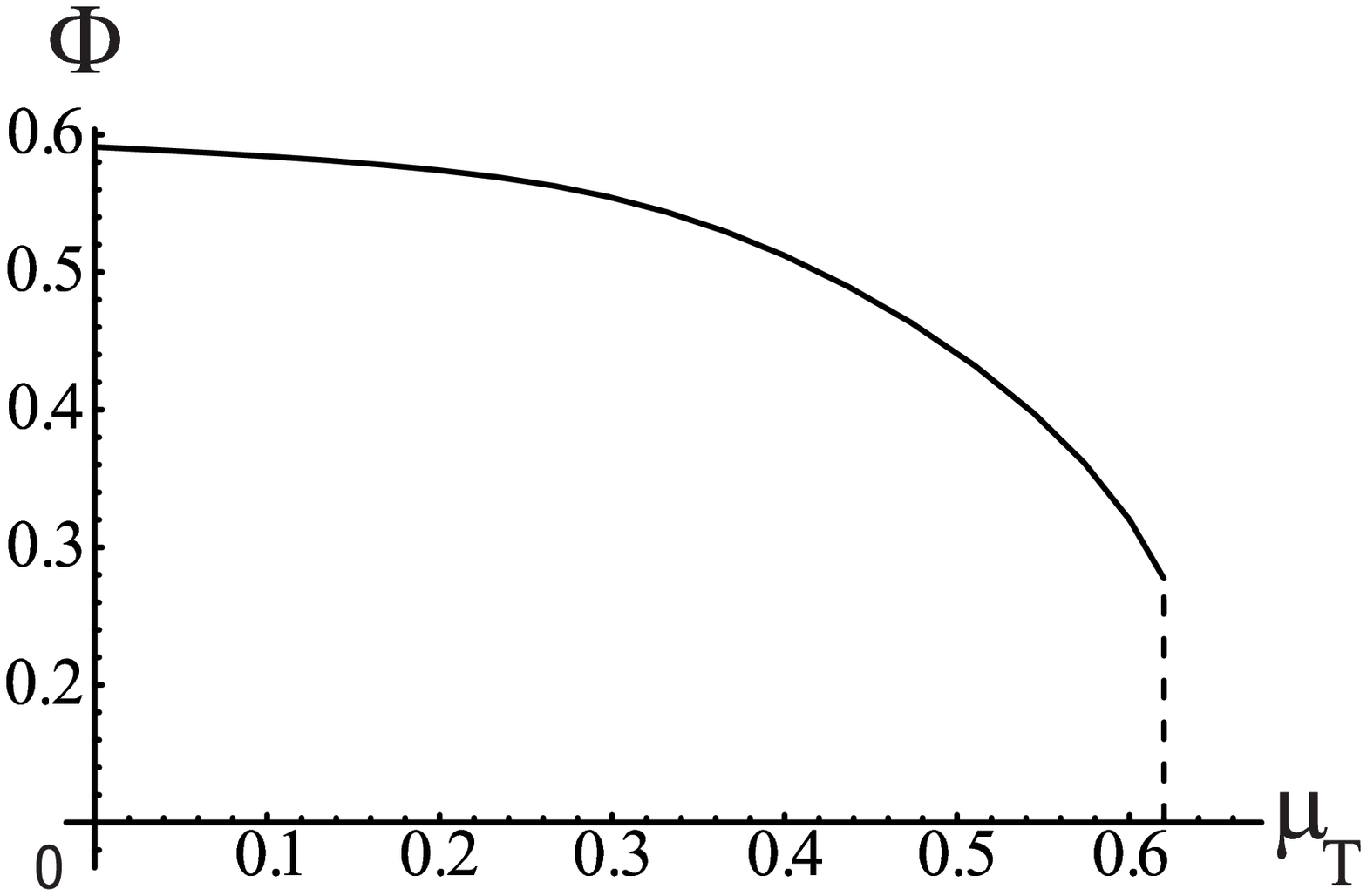}
\caption{Expectation value $\Phi$ as a function of the chemical
potential $\mu_T=-\mu_s+\mu'_s$ at $T=1.5<T_{\rm tri}$.
The phase transition is of first order.
}
\label{figPhi1.5}
\end{center}
\end{figure}

\begin{figure}[htbp]
\begin{center}
\epsfxsize=6cm
\epsffile{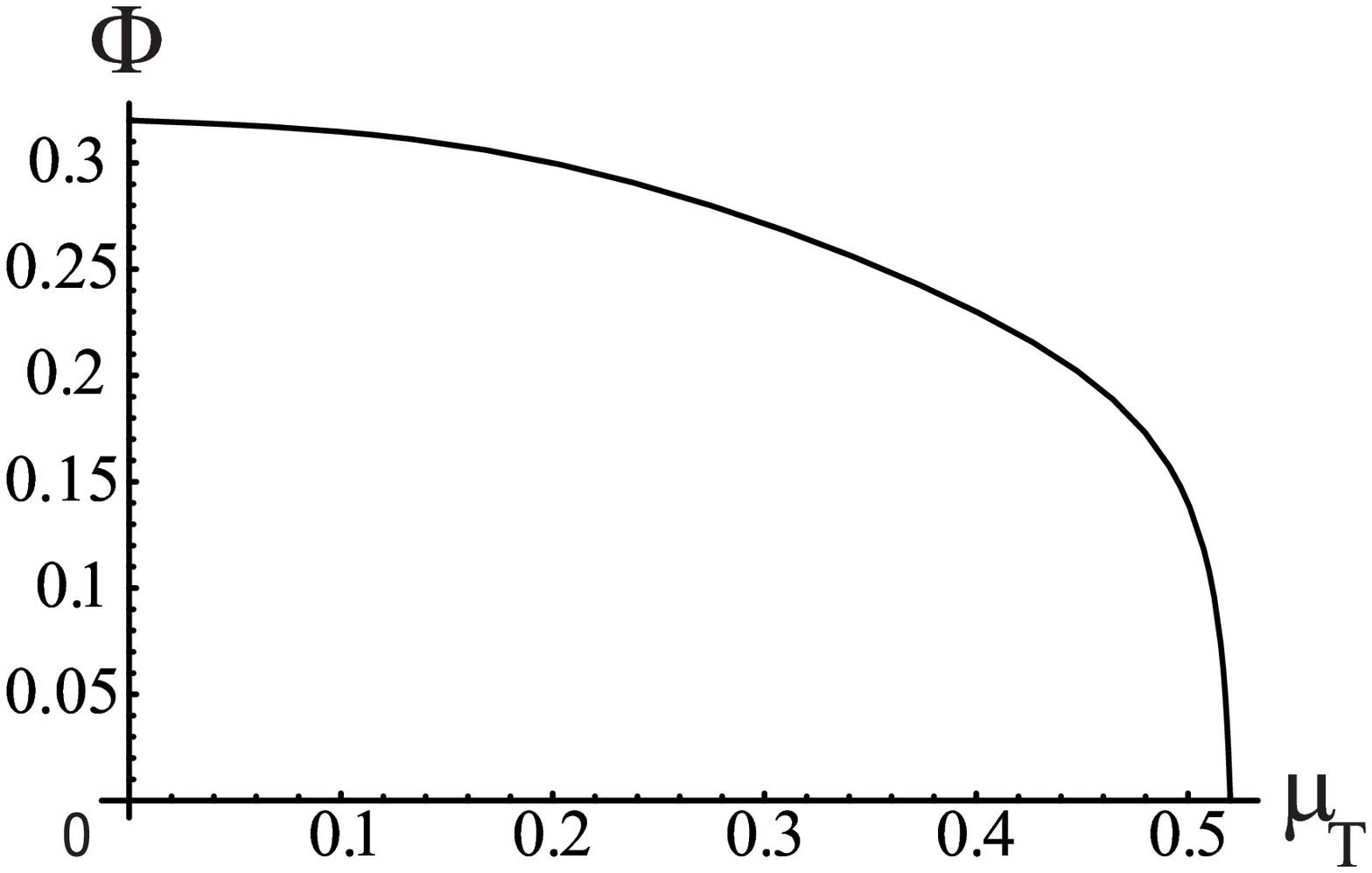}
\caption{Expectation value $\Phi$ as a function of the chemical
potential $\mu_T=-\mu_s+\mu'_s$ at $T=2.5>T_{\rm tri}$.
The phase transition is of second order.
}
\label{figPhi2.5}
\end{center}
\end{figure}

\begin{figure}[htbp]
\begin{center}
\epsfxsize=6cm
\epsffile{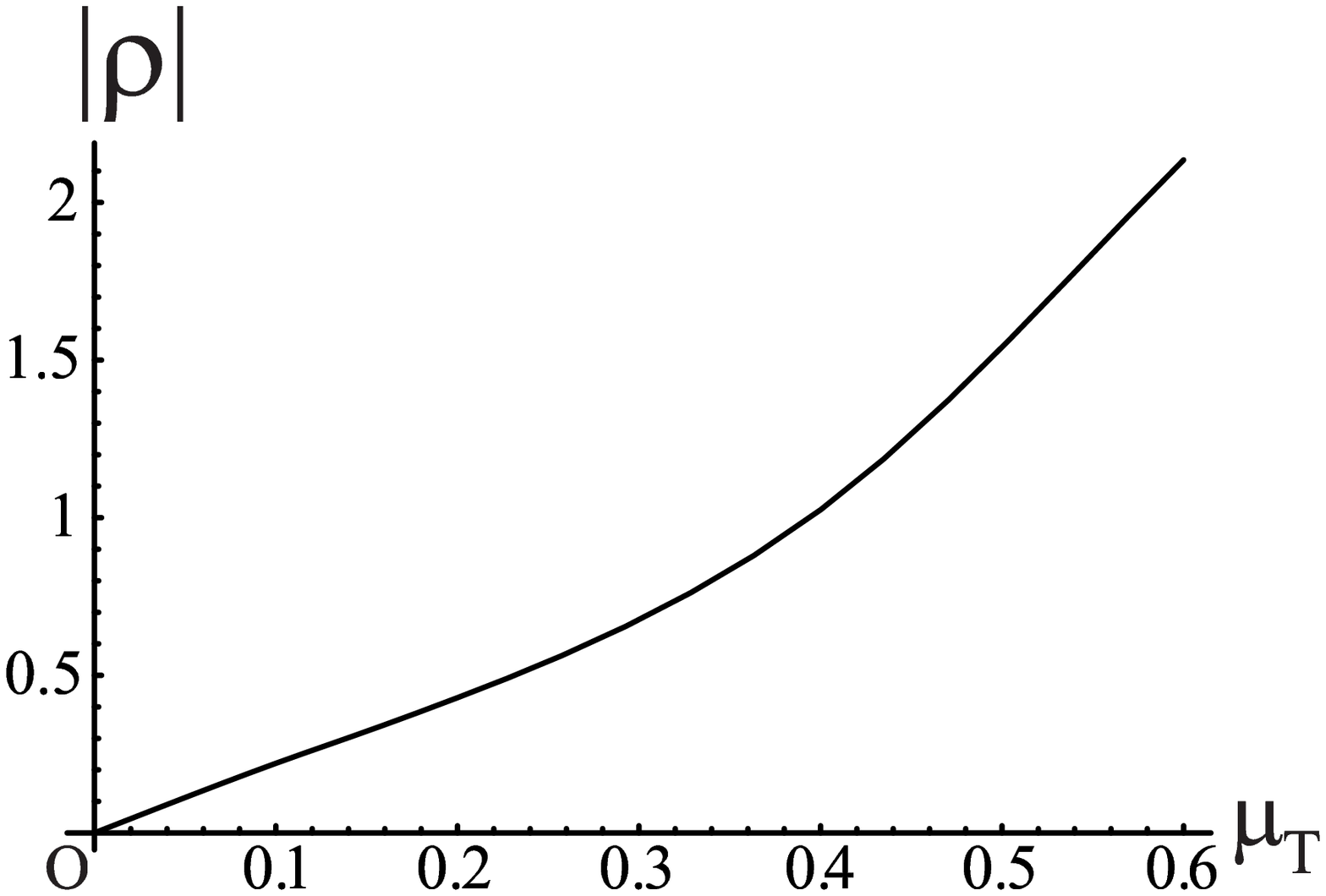}
\caption{Density of spinons $\rho(<0)$ as a function of the chemical
potential $\mu_T=-\mu_s+\mu'_s$ at $T=1.5<T_{\rm tri}$.
}
\label{figrho1.5}
\end{center}
\end{figure}

\begin{figure}[htbp]
\begin{center}
\epsfxsize=6cm
\epsffile{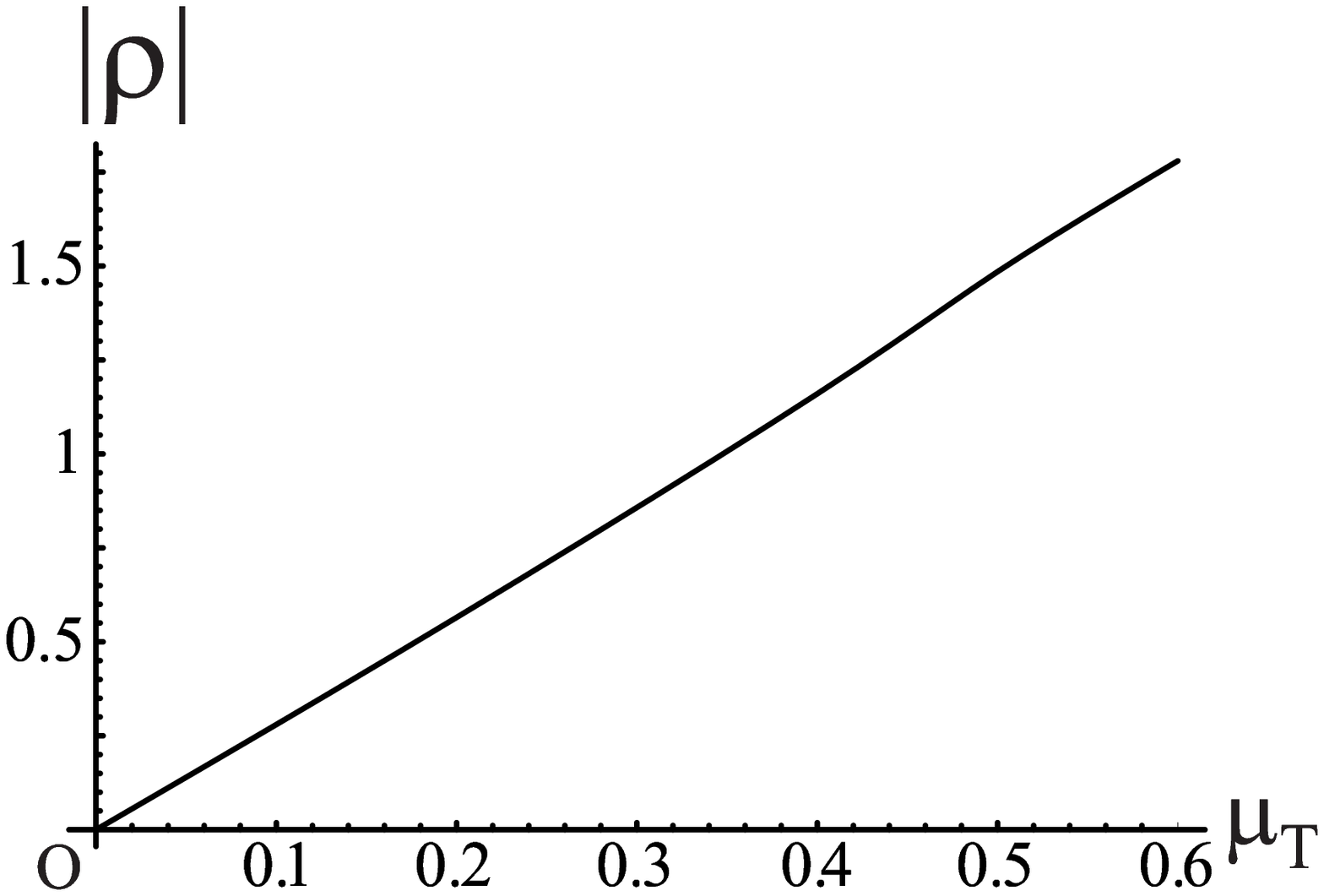}
\caption{Density of spinons $\rho(<0)$ as a function of the chemical
potential $\mu_T=-\mu_s+\mu'_s$ at $T=2.5>T_{\rm tri}$.
}
\label{figrho2.5}
\end{center}
\end{figure}

Finally let us show numerical calculations for some typical
values of the parameters in $F_{\rm eff}$.
We put $d=3, \; B=1,\; C=0$ and $N_f=4$ in Eq.(\ref{Fef2}).
In this case, $T_N=3,\; T_{\rm tri}=2,\; \mu_T^{\rm tri}=0.658\cdots,\;
\mu^c_{T0}=0.549\cdots$ and $\Phi_0=0.693\cdots$.
In Fig.\ref{figPD}, we show the phase diagram in the $\mu_T-T$ plane.
The spinon-gauge system at the strong-coupling limit has the symmetry-breaking
phase in the low-$T$ and low-density region.
In the high-$T$ region $T>T_{\rm tri}$, the phase transition is of
second-order whereas in the low-$T$ region $T<T_{\rm tri}$ it is of first-order.At quantum critical point $(\mu_T,T)=(\mu^c_{T0},0)$, the density of spinons 
changed drastically from zero to $-N_f$.
The phase transition line show a reentrant behavior, but this behavior
may be an artifact of the mean-field approximation.
See discussion in the following section.
In Figs.\ref{figPhi1.5} and \ref{figPhi2.5}, we show the expectation value of
$\Phi$ as a function of $\mu_T$.
For $T<T_{\rm tri}$, $\Phi$ exhibits a discontinuity at $\mu_T=\mu^c_{T0}$.
In Figs.\ref{figrho1.5} and \ref{figrho2.5}, we show the density 
$\rho$ as a function of $\mu_T$.

\setcounter{equation}{0}
\section{Discussion and conclusion}

In this paper we studied the strong-coupling region of the spinon-gauge 
system on the lattice at finite-$T$ and finite density of spinons.
By using the MFT like approximation, we found that at low-$T$ and low
density the system has the AF long-range order (LRO).
We expect that this LRO is weak magnetism compared with the 
microscopic local N\'eel order as explained in Ref.\cite{SVS} and 
observed by the experiments of the heavy-fermion systems. 
Low-energy excitations in the ordered phase are easily identified as follows 
by the symmetry of the system Eq.(\ref{sym2}) and also from the result of 
the one-link integral Eq.(\ref{onelink2}),
\begin{equation}
\Phi_x=\langle \Phi\rangle\cdot \exp \Big[i\epsilon(x)\sum_KT^K\pi^K(x)\Big],
\label{pi}
\end{equation}
where $\epsilon(x)=(-)^{x_0+\cdots+x_{d-1}}$, $T^K$ is the generators
of U($N_f$) and $\pi^K(x)$ are gapless excitations, the spinwaves. 
Then one can easily expect that an effective field theory for the ordered 
phase is a U($N_f$) nonlinear $\sigma$-model.

At higher-$T$, the effect of fluctuations of $\pi^K(x)$ is larger
and destabilizes the AF order.
Then the reentrance of the transition line observed in the MFT like
approximation may disappear as a result of the fluctuations 
described by $\pi^K(x)$.

As discussed rather in detail in Ref.\cite{MF}, the MFT-like approximation
is reliable for small $N_f$, whereas for large $N_f$ 
an instability of the state with long-range order to a symmetric state appears.
Therefore the strong-coupling gauge theory and also the  MFT-like approximation
employed in the present paper is justified only for small $N_f$ cases.

In this paper we considered the gauge theory of spinons.
It is quite interesting to study a coupled system of the conduction
electrons and spin degrees of freedom and discuss the quantum phase
transition to heavy-fermion state\cite{SVS}.


\end{document}